\PassOptionsToPackage{unicode}{hyperref}
\PassOptionsToPackage{hyphens}{url}
\documentclass[
]{article}
\usepackage{amsmath,amssymb}
\usepackage{iftex}
\ifPDFTeX
  \usepackage[T1]{fontenc}
  \usepackage[utf8]{inputenc}
  \usepackage{textcomp} 
\else 
  \usepackage{unicode-math} 
  \defaultfontfeatures{Scale=MatchLowercase}
  \defaultfontfeatures[\rmfamily]{Ligatures=TeX,Scale=1}
\fi
\usepackage{lmodern}
\ifPDFTeX\else
\fi
\IfFileExists{upquote.sty}{\usepackage{upquote}}{}
\IfFileExists{microtype.sty}{
  \usepackage[]{microtype}
  \UseMicrotypeSet[protrusion]{basicmath} 
}{}
\makeatletter
\@ifundefined{KOMAClassName}{
  \IfFileExists{parskip.sty}{%
    \usepackage{parskip}
  }{
    \setlength{\parindent}{0pt}
    \setlength{\parskip}{6pt plus 2pt minus 1pt}}
}{
  \KOMAoptions{parskip=half}}
\makeatother
\usepackage{xcolor}
\usepackage[margin=1in]{geometry}
\usepackage{longtable,booktabs,array}
\usepackage{calc} 
\usepackage{etoolbox}
\makeatletter
\patchcmd\longtable{\par}{\if@noskipsec\mbox{}\fi\par}{}{}
\makeatother
\IfFileExists{footnotehyper.sty}{\usepackage{footnotehyper}}{\usepackage{footnote}}
\makesavenoteenv{longtable}
\usepackage{graphicx}
\makeatletter
\def\maxwidth{\ifdim\Gin@nat@width>\linewidth\linewidth\else\Gin@nat@width\fi}
\def\maxheight{\ifdim\Gin@nat@height>\textheight\textheight\else\Gin@nat@height\fi}
\makeatother
\setkeys{Gin}{width=\maxwidth,height=\maxheight,keepaspectratio}
\makeatletter
\def\fps@figure{htbp}
\makeatother
\NewDocumentCommand\citeproctext{}{}

\makeatletter
 \let\@cite@ofmt\@firstofone
 \def\@biblabel#1{}
 \def\@cite#1#2{{#1\if@tempswa , #2\fi}}
\makeatother
\newlength{\cslhangindent}
\newlength{\csllabelwidth}
\newenvironment{CSLReferences}[2] 
 {\begin{list}{}{%
  \setlength{\itemindent}{0pt}
  \setlength{\leftmargin}{0pt}
  \setlength{\parsep}{0pt}
  \ifodd #1
   \setlength{\leftmargin}{\cslhangindent}
   \setlength{\itemindent}{-1\cslhangindent}
  \fi
  \setlength{\itemsep}{#2\baselineskip}}}
 {\end{list}}
\usepackage{calc}

\usepackage{setspace}
\usepackage{booktabs}
\usepackage{longtable}
\usepackage{array}
\usepackage{multirow}
\usepackage{wrapfig}
\usepackage{float}
\usepackage{colortbl}
\usepackage{pdflscape}
\usepackage{tabu}
\usepackage{threeparttable}
\usepackage{threeparttablex}
\usepackage[normalem]{ulem}
\usepackage{makecell}
\usepackage{xcolor}
\ifLuaTeX
  \usepackage{selnolig}  
\fi
\usepackage{bookmark}
\IfFileExists{xurl.sty}{\usepackage{xurl}}{} 
\hypersetup{
  pdftitle={The Fast and the Furious: Tracking the Effect of the Tomoa Skip on Speed Climbing},
  pdfauthor={Caleb Chou; Andee Kaplan},
  hidelinks,
  pdfcreator={LaTeX via pandoc}}

\title{The Fast and the Furious: Tracking the Effect of the Tomoa Skip on Speed Climbing}
\author{Caleb Chou\footnote{Department of Computer Science, Colorado State University} \and Andee Kaplan\footnote{Department of Statistics, Colorado State University}}
\date{}

\begin{document}
\maketitle

\section{Welcome to the World of Competitive Climbing}\label{welcome-to-the-world-of-competitive-climbing}

Rock climbing, a formerly niche sport, has grown rapidly since its emergence in the late 19th century. From the early days of alpine expeditions to the Olympic games, the sport of climbing has shifted and evolved as technologies and techniques are built upon and refined by the rapid influx of climbers continuously picking up the sport. In the United States alone there are estimated to be 10.35 million rock climbers as of 2021. What was once a pursuit of few has now become a pastime for millions, a unique blend of athleticism, mental-fortitude, and strategy unlike any other.

Competitive climbing can be broken up into three main disciplines: lead climbing, bouldering, and speed climbing. Governed by the International Federation of Sport Climbing (IFSC), climbing competitions are held for each distinct discipline. Lead climbing is a style of climbing in which climbers scale a climbing wall using a set of artificial holds. These climbing walls are generally long and require climbers to tie into a rope which they then use to clip into protection drilled into the wall. On the other hand, bouldering involves tackling short yet challenging sequences of holds on smaller walls. Due to the shorter height, boulderers use shock-absorbing mats to cushion their falls instead of relying on ropes. Although lead climbing and bouldering are completely different disciplines of climbing, many of the same fundamental techniques, physical strengths, and problem-solving abilities are needed to excel in both styles.

\begin{figure}

{\centering \includegraphics[width=0.5\textwidth]{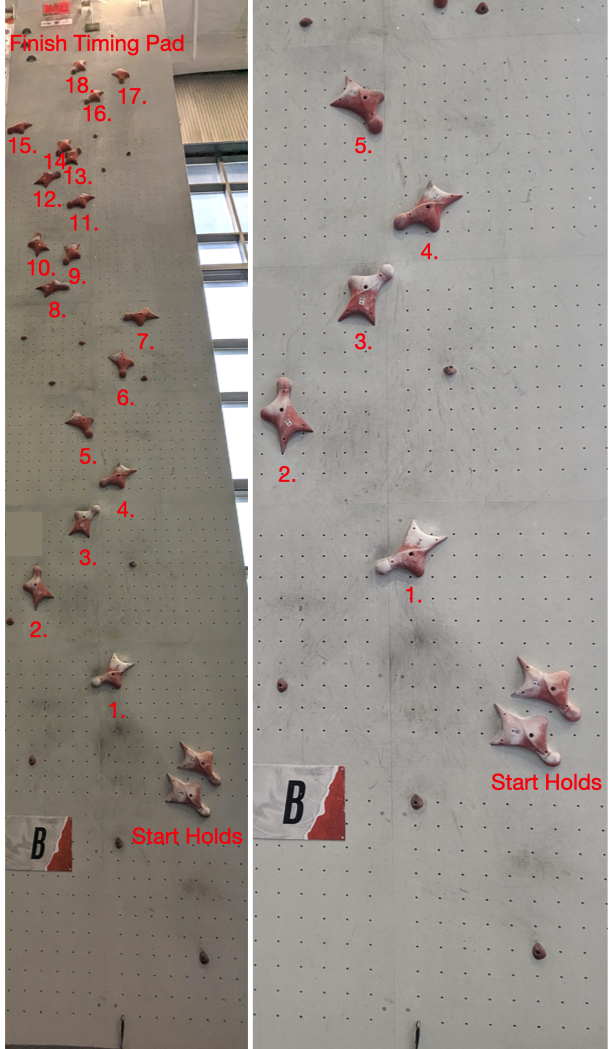} 

}

\caption{(LEFT) The speed wall layout with hand holds labeled in order. (RIGHT) A zoom in on the first five holds of the speed wall.}\label{fig:speed-wall}
\end{figure}

Competitive speed climbing is the newest discipline out of the three to gain popularity. This style of climbing involves scaling a fixed wall as fast as possible. The unique aspect of speed climbing is that the holds on the speed climbing wall are fixed. In other words, these holds are placed in exactly the same angle and position every single time the wall is climbed. The speed route is 15 meters long, sits at a 5 degree angle, and consists of 20 large hand holds and 11 smaller foot holds (shown in Figure \ref{fig:speed-wall}). This is vastly different from the variable nature of sport climbing and bouldering. Speed climbing demands flawless technique and exceptional power. Speed climbers must refine their movements in such a way that they are both extraordinarily fast and extremely consistent. In an interview conducted by GQ with professional competition climber Kyra Condie, she says of the importance of consistency in speed climbing, ``Having that muscle memory makes you go faster because hitting everything perfectly is one thing that takes off a fraction of a second.'' This muscle memory allows climbers to fly up the 15 meter speed wall relying almost purely on instinct and muscle memory. The best speed climbers are able to complete the route in under 5 seconds.

IFSC speed climbing competitions consist of two rounds: a qualifying and a final. In the qualifying round, climbers have two attempts to score their best time by climbing the speed wall as fast as possible. The climbers with the top 16 qualifying times then advance to the final round where they compete in head to head races against other qualifiers. The winner of each of these head to head races then advances to the next round of the final. Consequently, the amount of competitors in the final round are cut in half through each round of head to head races until there is a winner. This format necessitates that athletes climb extremely fast in the qualifying round while also maintaining consistent speeds throughout the final rounds.

The disparities between speed climbing and lead climbing or bouldering are so substantial that climbers generally tend to specialize either in lead climbing and/or bouldering, or solely in speed climbing. Lead climbing and bouldering typically focus on strength, endurance, power, and ``route finding'', or the aspect of puzzling out how a climber can move through the climbing holds. On the other hand, speed climbing focuses on rapid bursts of power and the ability to fine-tune movements to remain consistently fast on the wall. These differences in skill sets between disciplines means that climbers must train differently based on their specialty. In a 2016 survey conducted by Climbing Magazine, one climber noted of speed climbing, ``Those skills/abilities don't transfer as well to the other disciplines.'' Consequently the climbing community was left in shock when it was announced that in the 2020 Olympics climbers would have to participate in all three disciplines. In a Yahoo!Sports article, one climber is quoted as saying, ``It's like getting the track athletes that have always just been single discipline athletes to suddenly do the shot put, the 100 meters and the 1500 or something.'' Climbers from every discipline of the sport were upset by the decision to format the event in such a manner. Despite confusion surrounding the event format, these athletes would have to quickly learn a completely new set of skills if they hoped to compete in the upcoming Olympic event.

\begin{figure}
\includegraphics[width=\textwidth]{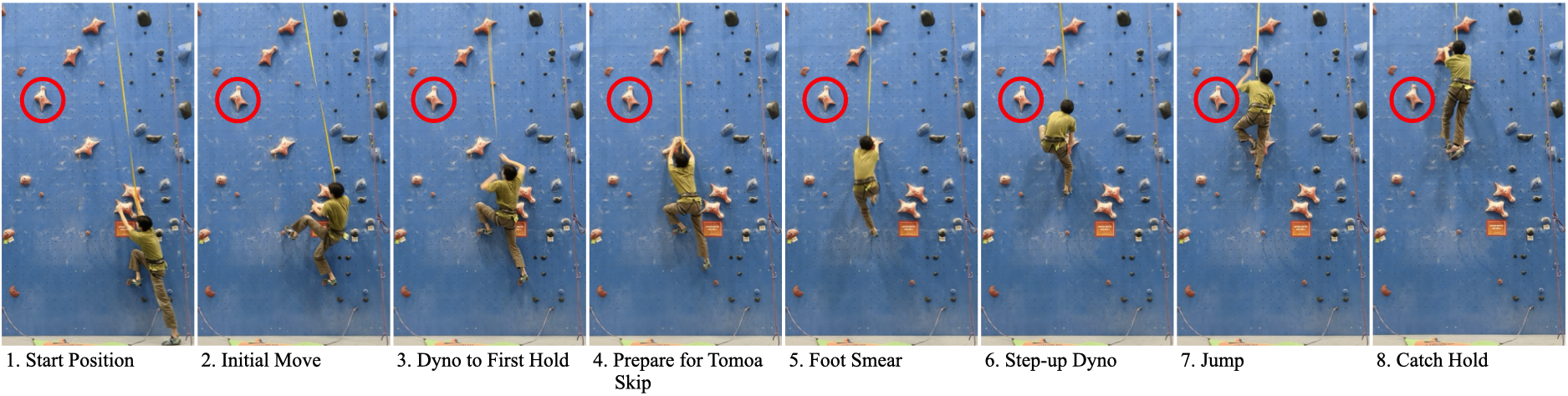} \caption{Demonstration of the initial sequence of the speed route by the author.  The Tomoa Skip comprises moves 5-8 in the sequence of figures. The hold that is being skipped is circled in red.}\label{fig:ts-demo}
\end{figure}

Thus, in preparation for the 2020 Olympic games there was a surge of boulderers and lead climbers scrambling to pick up speed climbing. One of these climbers, Tomoa Narasaki, was a competitive boulderer hailing from Japan. Like many other athletes, Narasaki had entered the foreign world of speed climbing in preparation for the Olympic games. Drawing upon his bouldering expertise, Narasaki quickly introduced a groundbreaking speed climbing technique later dubbed the Tomoa Skip. This new technique blended the powerful and dynamic style of bouldering with the rapid ascent of speed climbing. The Tomoa Skip is a strategic approach, often referred to as a ``beta'', to climbing through the initial segment of the speed climbing wall (see the right side of Figure \ref{fig:speed-wall}). This approach centers around the first five handholds of the speed wall and involves climbers propelling themselves up and over a particular handhold, allowing them to bypass the utilization of a hold off to the left. This crucial sequence involves a move known as a ``step-up dyno''. A dyno is a climbing move that involves very dynamic movement, typically relying on a climber using their momentum to throw themselves in some direction. The step-up dyno is a special type of dyno that involves a climber stepping up onto a hold that they are currently grabbing and then propelling themselves upwards over the hold. The Tomoa Skip is so named because the move requires a step-up dyno to skip hold number 2. on the speed wall. The climber uses the step-up dyno on hold number 1. to move directly to hold number 3. A demonstration of the Tomoa Skip by the author is included in Figure \ref{fig:ts-demo}.

The introduction of the Tomoa Skip marked a shift in the speed climbing community. The fusion of bouldering and speed climbing techniques sparked by the upcoming Olympic games brought forth a completely new dimension of technique and strategy that was quickly adopted by the community as a whole. This paper aims to delve into the impact of the Tomoa Skip on speed climbing times and explore the possibility that this groundbreaking technique has led to increased variability in climbing times due to the potentially riskier movement. We next discuss the data collection techniques used before answering the questions, ``Are climbers who use the Tomoa Skip faster?'' and ``Are these climbers less consistent?''

\section{Data Collection and Initial Trends}\label{data-collection-and-initial-trends}

To answer these questions, we turned to the catalog of speed climbing records maintained by the IFSC. This repository encompasses speed climbing competition results dating back as far as 1990 and extends up to the present date. To retrieve these results we made use of the Selenium automated web browser to develop a web scraper that extracted information from HTML tables. This web scraper targeted the result pages from speed climbing competitions held from 2012 to 2022, a period chosen because speed climbing result data is most prevalent from 2012 onward and data recorded prior to 2012 is sparser and less consistent to scrape. To do this, the scraper iterates through each event listed on the IFSC website, capturing the unique ID of each of these events as well as their name and start and end dates. From here, the scraper iterates through these unique event IDs corresponding to events in our chosen date ranges and constructs a URL used to navigate to the competition result page corresponding to each event. Once this page has been navigated to, the scraper retrieves and stores the events data using the \texttt{rvest} package to transform an HTML table into an R dataframe. From this process, we gather information pertaining to individual climber's performance in the qualifying and final rounds of each speed climbing competition that they have competed in.

After collecting this competition result data, it was necessary to determine which climbers utilized the Tomoa Skip and in which events they used it in. As this data is not recorded anywhere, the only way to determine which climbers used the Tomoa Skip was to view videos of climbing competitions broadcasted by the IFSC and record which individual climbers used the Tomoa Skip in each competition. This was done by viewing filmed broadcasts of the final rounds of speed climbing competitions held from 2018 to 2022 on the IFSC YouTube channel. This specific date range was chosen because Tomoa Narasaki did not introduce the Tomoa Skip until 2018. In order to determine which climbers employed the Tomoa Skip, we slowed each video down to half speed or slower, giving us a more detailed view of each climber's movements. For each climber competing in the video we were viewing, we watched for whether they executed the Tomoa Skip by identifying whether they used their feet on the first hold to propel themselves up to the third hold, thus performing a step up dyno. Through this process, we examined 54 videos which amounts to over 108 hours of footage.

Although our web scraper retrieved data for qualifying and final rounds, the IFSC only broadcasts the final rounds of speed climbing competitions. This necessitated the creation of a script that implemented our assumption that if a climber uses the Tomoa Skip in one round of an event, they will continue to use it in the future. This approach enables us to bridge the gaps in our dataset. For instance, if we have a documented occurrence of a climber using the Tomoa Skip in one event, looking ahead, if this climber doesn't progress to a final round in future competitions we can still maintain a record indicating whether or not they utilized the skip. In order to avoid overgeneralizing, the script developed to implement this assumption takes into account the fact that some climbers may switch from using the Tomoa Skip to not using the Tomoa Skip. In this case, we simply perform the reverse action and assume that this climber will continue to not use the Tomoa Skip in any future competitions up until we have a record indicating otherwise. This case, however, is extremely rare, with only one climber in our records documented as having made this switch back.

In order to investigate the effect of a climber's age on their speed climbing times, we went through another manual data collection process. We constructed a comprehensive spreadsheet documenting each distinct climber, totaling 952 athletes. This spreadsheet includes the climber's date of birth, the source from which the date of birth was obtained, and the date on which that source was accessed. For climbers whose date of birth was only available as a year, we approximated their date of birth as January 1st of that year. Additionally, for certain climbers for whom only age information was available, we approximated their date of birth by subtracting their age from the current year. Once this data collection was complete, we ran a script to determine each climbers age at the time of each event that they competed in. This was achieved by subtracting the climber's date of birth from the event's start date. To address the issue of missing values, which amounted to approximately 100 instances, we filled in the climber's age with the average age of athletes within the respective event. Additionally, we implement a script to create a ``time progression'' variable within our dataset. This variable calculates the number of days since the first observed competition, providing a temporal context as well as sequencing to different events.

\begin{figure}

{\centering \includegraphics[width=1\linewidth]{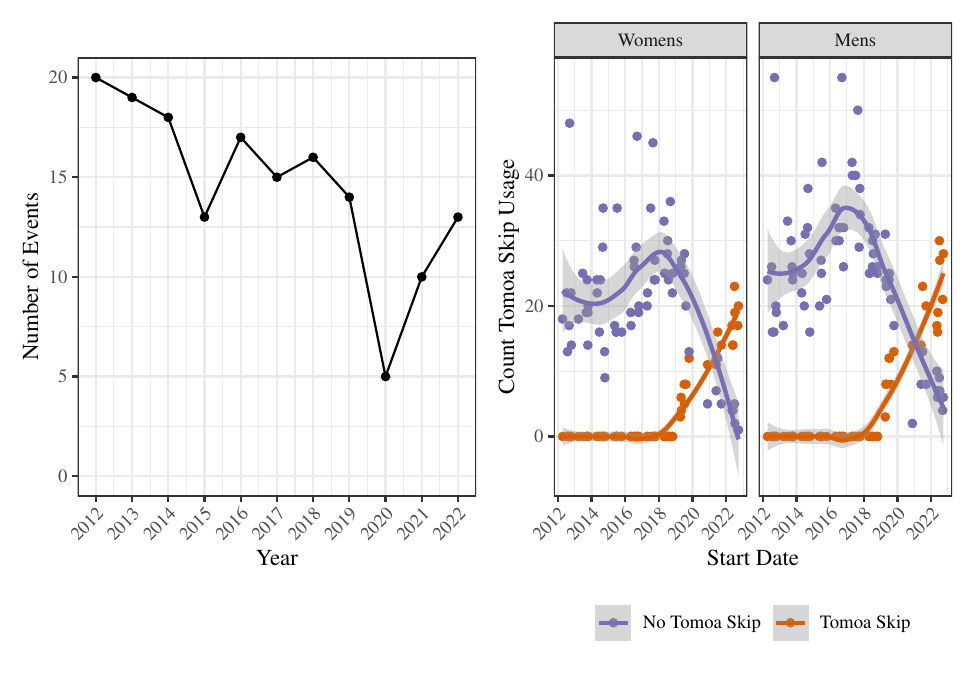} 

}

\caption{(LEFT) Number of events per year. There is an average of approximately 15 events per year, with a notable deviation in the year 2020 due to COVID. (RIGHT) Tomoa Skip usage over time. Since the introduction of the move in 2018 there has been a steady increase in Tomoa Skip usage for both men and women. Each point in the plot indicates a single IFSC event, colored by the number of climbers in attendance that used the Tomoa Skip. The overlayed lines are a LOESS smoother, showing the trends over time.}\label{fig:events-per-year}
\end{figure}

From the data we gathered, we can ascertain that between the years 2012 and 2022, there is an average of around 15 major speed climbing competitions that are held in each year. The number of events per year is shown on the left side of Figure \ref{fig:events-per-year}. Notably, in the year 2020, this pattern deviates, with only five competitions held. This can be attributed to the disruptive impact of COVID, as global lockdowns significantly impeded the ability of athletes to participate and compete in climbing competitions. In addition to this, we found that the speed climbing community quickly embraced the Tomoa Skip after its introduction in 2018. By 2022, the Tomoa Skip had become the predominant choice among professional speed climbers, and by the end of that year, there are few climbers remaining who do not use the skip (see the right side of Figure \ref{fig:events-per-year} for reference).

\begin{table}[H]
\centering
\caption{\label{tab:fall-false-starts}Fall and False Start Rates in Final Rounds for Men and Women}
\centering
\begin{tabular}[t]{lrrrrr}
\toprule
Category & Events & Falls & False Starts & Fall Rate & False Start Rate\\
\midrule
Women & 69 & 105 & 24 & 1.52 & 0.35\\
Men & 69 & 141 & 57 & 2.04 & 0.83\\
\bottomrule
\end{tabular}
\end{table}

Our data offers insight into the performance of 952 distinct climbers during this time span, composed of 549 male climbers and 403 female climbers. Through our exploratory analysis we found that across 69 events, men experienced a fall rate of 2.04 during final rounds whereas women experienced a slightly lower fall rate of 1.52. This signifies that in a final round of a competition, on average, approximately 2 out of 16 male climbers will fall off of the wall. Similarly, men had a false-start rate of 0.83 during final rounds, whereas women had a false-start rate of 0.35. This translates to an average of 0.83 out of 16 male climbers starting their climb prematurely, resulting in their disqualification from the round. These fall and false start rates can be seen in Table \ref{tab:fall-false-starts}.

\section{How Important is the Tomoa Skip?}\label{how-important-is-the-tomoa-skip}

We now turn to answering the question of the importance of the Tomoa Skip on speed climbing.

\subsection{Effect of the Tomoa Skip on Speed Times}\label{effect-of-the-tomoa-skip-on-speed-times}

In order to determine whether the Tomoa Skip has significantly impacted speed climbing times, we made use of mixed effects models. Mixed models have an extensive history of being applied in sports research to understand the effects of various factors on athletic performance. See the further reading section for more details. These models allow us to account for the inherent variability between climbers as well as to address the non-independence of individual climber performance. This is particularly valuable given that climbers exhibit a diverse range of skill levels and each may undergo differing rates of improvement. Additionally, a climber's performance in one event may be influenced by their performance in a prior event, a fact that would violate the assumption of independence in the traditional linear regression model. This is accounted for by random effects, which allow us to consider variations in both the initial skill level and the rate of progression of individual climbers. Additionally, we can examine the fixed effects of a climber's gender, age, and time progression (a variable that provides both a temporal context and a sequencing for different events), in order to determine whether these factors help us better understand individual athletes' performances in speed climbing. If we let \(x_{1ij}\) denote use of the Tomoa Skip at the \(i^\text{th}\) event by the \(j^\text{th}\) climber, \(x_{2j}\) denote the gender of the \(j^\text{th}\) climber, \(x_{3ij}\) denote the age of the \(j^\text{th}\) climber at the \(i^\text{th}\) event, \(x_{4i}\) denote the time progression at the \(i^\text{th}\) event, and \(y_{ij}\) denote climber \(j\)'s best time at the \(i^\text{th}\) event, then we can formalize a sequence of potential models as follows:

\begin{align*}
\mathcal{M}_0&: \parbox[t]{0.75\textwidth}{A random effect for climbers and a random intercept for each event} \\
\mathcal{M}_1&: \parbox[t]{0.75\textwidth}{Random effects for both climbers and events} \\
\mathcal{M}_2&: \parbox[t]{0.75\textwidth}{A fixed effect for gender and random effects for both climbers and events} \\
\mathcal{M}_3&: \parbox[t]{0.75\textwidth}{A fixed effect for Tomoa Skip usage, gender, age, and time progression and random effects for climbers and events} \\
\mathcal{M}_4&: \parbox[t]{0.75\textwidth}{An interaction between Tomoa Skip usage and age, a fixed effect for gender and time progression, and random effects for climbers and events.}
\end{align*}

In particular, we can formalize \(\mathcal{M}_3\) using previously defined notation as follows:
\begin{align*}
\log(y_{ij}) &= \beta_{0ij} + \beta_{1ij} x_{1ij} + \gamma_{02} x_{2j} + \gamma_{03} x_{3ij} +\gamma_{04} x_{4i} + \epsilon_{ij}  \\
\beta_{0ij} &= \gamma_{00} + \mu_{0i} + \upsilon_{0j} \\
\beta_{1ij} &= \gamma_{01} + \mu_{1i} + \upsilon_{1j} \\ 
\begin{pmatrix} \mu_{0i} \\ \mu_{1i} \end{pmatrix} &\stackrel{\text{iid}}{\sim} \text{MVN}_2 
    \begin{pmatrix} \begin{pmatrix} 0 \\ 0 \end{pmatrix} , 
    \begin{pmatrix} \tau_{00} & \tau_{01} \\ \tau_{01} & \tau_{11}\end{pmatrix} 
\end{pmatrix} \\ 
\begin{pmatrix} \upsilon_{0j} \\ \upsilon_{1j} \end{pmatrix} &\stackrel{\text{iid}}{\sim} \text{MVN}_2 
    \begin{pmatrix} \begin{pmatrix} 0 \\ 0 \end{pmatrix} , 
    \begin{pmatrix} \eta_{00} & \eta_{01} \\ \eta_{01} & \eta_{11}\end{pmatrix} 
\end{pmatrix} \\
\epsilon_{ij} &\stackrel{\text{iid}}{\sim}\text{N}(0, \sigma^2).
\end{align*}

\begin{table}

\caption{\label{tab:model-comparison-anova}ANOVA table output of our model comparisons.}
\centering
\begin{tabular}[t]{lccl}
\toprule
Models & BIC & Statistic & p-value\\
\midrule
$\mathcal{M}_{0}$ & -3502.916 & NA & NA\\
$\mathcal{M}_{1}$ & -5871.640 & 2409.6545 & 0.0000\\
$\mathcal{M}_{2}$ & -6053.375 & 189.9215 & 0.0000\\
\textbf{$\mathbf{\mathcal{M}}_{3}$} & \textbf{-6126.498} & \textbf{97.6818} & \textbf{0.0000}\\
$\mathcal{M}_{4}$ & -6122.896 & 4.5838 & 0.0323\\
\bottomrule
\end{tabular}
\end{table}

In addition to this proposed model, we compared models of various complexities in order to determine which combination of fixed and random effects best models the effect of the Tomoa Skip on speed climbing times. The models we selected for comparison contain varying degrees of complexity and give us a diverse range of options for evaluation. To evaluate these models we can compare the BIC as well as perform an ANOVA to test if the sequence of increasingly complex models is significantly better at capturing the relationships present in the data than the simpler model. The results of these two forms of evaluation are presented in Table \ref{tab:model-comparison-anova}. If we select our model based off of the BIC we aim for the model with the lowest value. Similarly, the ANOVA test tells us whether or not it is statistically beneficial to add more terms. Based on the results of these comparisons, both BIC and ANOVA with \(\alpha = 0.01\) indicate that we select \(\mathcal{M}_3\) as our best chosen model.

In the constructed model, we use the logarithm of a climber's best time for each event as the outcome variable due to the right-tailed nature of best times. This best time is determined by taking the climber's quickest performance out of all the rounds of a given event. This means that for every event a climber competes in, our model will only consider the round of that event in which they climbed the fastest. Applying the logarithm to these best times helps us to align more effectively with the assumptions of our model. For predictors, we use an indicator of the climbers gender as a standard fixed effect and allow for a random effect in the intercept and slope associated with use of the Tomoa Skip that varies by both climber and event. In using a fixed effect for climber's gender we are able to account for how differences in gender may affect times. In using random effects that allow the intercept and slope associated with use of the Tomoa Skip to vary with both climber and event, we can account for variability between climbers as well as between events. For example, an individual climber has a certain talent level and training regimen that may differ from another climber. Additionally, weather or wall conditions can impact the speed times at different events in different ways. For instance, a competition on an extremely rainy day may have a significant negative impact on the abilities of competing athletes. On the flip side, ideal temperatures and newer climbing holds could improve the grip of all climbers, potentially leading to faster times. In addition to this, we add a fixed effect for the age of a climber during an event and a fixed effect for time progression. The age fixed effect will account for variation in performance attributable to age. The time progression fixed effect enables us to control for temporal influences on speed climbing times, such as the overall improvement in performance trends over time.

The key differences between the assumptions of the standard linear regression model and the mixed effects model are the assumptions of independence and the distributions of the random effects. In the mixed effects model, we do not have to assume that observations within the same group are independent, however, each group must be independent of each other. In our case, this means that how one climber performs should not influence the performance of another climber. With the introduction of random effects, we now assume that the estimates of the random effects and residuals share an identical distribution.

\begin{table}

\caption{\label{tab:model-summary}Summary of our chosen mixed effect model's ($\mathcal{M}_3$) estimated fixed and random effects}
\centering
\begin{tabular}[t]{lcc}
\toprule
Term & Estimate & 95\% CI\\
\midrule
\addlinespace[.8em]
\multicolumn{3}{l}{\textbf{Fixed Effects}}\\
\hspace{1em}$\gamma_{00}$ & 2.0087 & (1.9368, 2.0806)\\
\hspace{1em}$\gamma_{01}$ & -0.1568 & (-0.1898, -0.1238)\\
\hspace{1em}$\gamma_{02}$ & 0.2830 & (0.2553, 0.3107)\\
\hspace{1em}$\gamma_{03}$ & -0.0931 & (-0.1128, -0.0733)\\
\hspace{1em}$\gamma_{04}$ & 0.0057 & (0.0028, 0.0085)\\
\addlinespace[.8em]
\multicolumn{3}{l}{\textbf{Random Effects}}\\
\hspace{1em}$\sigma^2$ & 0.0055 & \\
\hspace{1em}$\eta_{00}$ & 0.0736 & \\
\hspace{1em}$\tau_{00}$ & 0.0092 & \\
\hspace{1em}$\eta_{11}$ & 0.0286 & \\
\hspace{1em}$\tau_{11}$ & 0.0048 & \\
\hspace{1em}$\eta_{01}$ & -0.9808 & \\
\hspace{1em}$\tau_{01}$ & -0.9895 & \\
\bottomrule
\end{tabular}
\end{table}

The results of fitting this model are presented in Table \ref{tab:model-summary}. We observe an estimated fixed effect for the Tomoa Skip of -0.1568 with corresponding 95\% CI of (-0.1898, -0.1238). Because we took the logarithm of the climbers' best times in each event, this estimate signifies that after controlling for variation in individual climbers and events, using the Tomoa Skip is associated with an approximately 0.8549 times decrease in the average best climbing time (95\% CI: (0.8271, 0.8836)). For instance, if a climber has a best climbing time of 7 seconds without using the Tomoa Skip, we estimate that their best time would drop to approximately 5.9842 seconds using the Tomoa Skip. In the world of speed climbing, where a fraction of a second can mean the difference between success and failure, this seemingly small improvement constitutes a significant and impactful advancement.

\subsection{Effect of the Tomoa Skip on Variability of Speed Climbing Times}\label{variability}

\begin{figure}

{\centering \includegraphics[width=0.7\linewidth]{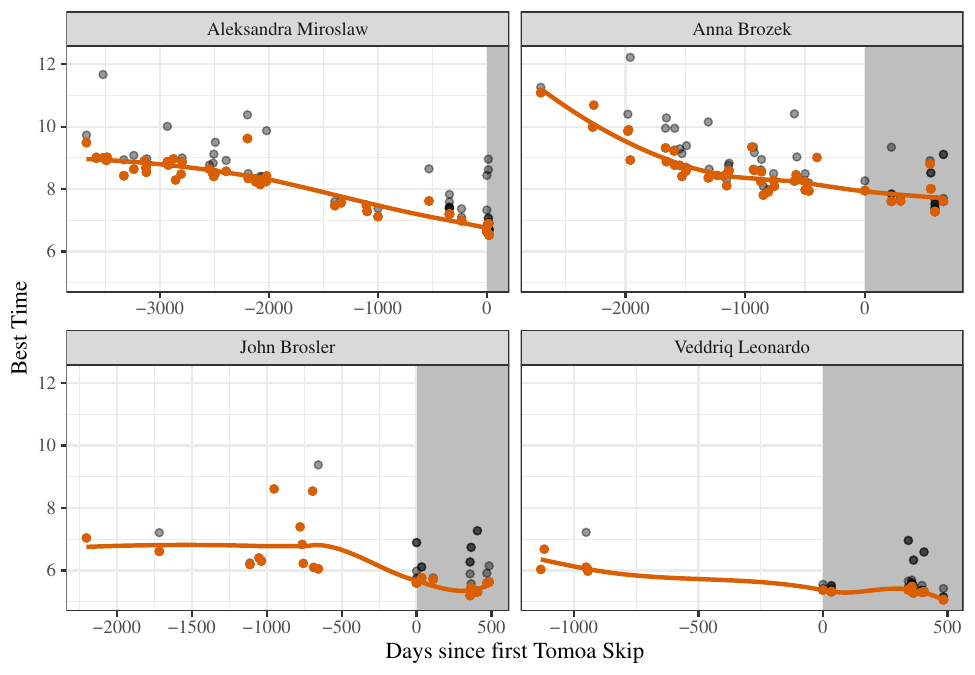} 

}

\caption{Progression of four climbers speed climbing times before and after adopting the Tomoa Skip. The grey rectangle denotes days after each climber has adopted the Tomoa Skip. Black points denote the climbers times in an event and red points denotes the climbers best time for each event. The lines are smoothed fits through the climbers best times to help show trends.}\label{fig:ts-progression}
\end{figure}

Although we have established that the Tomoa Skip has had a significant impact on speed climbing times, we now want to determine whether or not this improvement stays consistent. In other words, we want to determine whether or not adopting the Tomoa Skip has heightened the variability of speed climbing times. In order to investigate this question we can first examine the performance of several individual prominent speed climbers before and after they adopt the Tomoa Skip. Noticeably, speed climbers John Brosler and, to a lesser extent, Veddriq Leonardo both exhibit some form of increased variability in their climbing times after they adopt the Tomoa Skip. In other words, after adopting the Tomoa Skip, these climbers seem to have a wider range of climbing times. Although they achieve new personal records, and in the case of Veddriq Leonardo, world records, after adopting the Tomoa Skip, they also log a potentially unusual slew of slower times. Climber Anna Brozek appears to exhibit some form of variability in her climbing times after adopting the Tomoa Skip as well, recording her personal best time along with a scattering of slower times. While Alexsandra Miroslaw does not have as many recorded data points while using the Tomoa Skip as these other climbers, she exhibits a pattern of variability similar to that of the other climbers, recording personal bests (and world records) whilst also logging a plethora of slower times post adoption. Visually this can be seen in Figure \ref{fig:ts-progression} where the grey rectangle represents progressing days after the Tomoa Skip has been adopted by these climbers.

\begin{figure}

{\centering \includegraphics[width=1\linewidth]{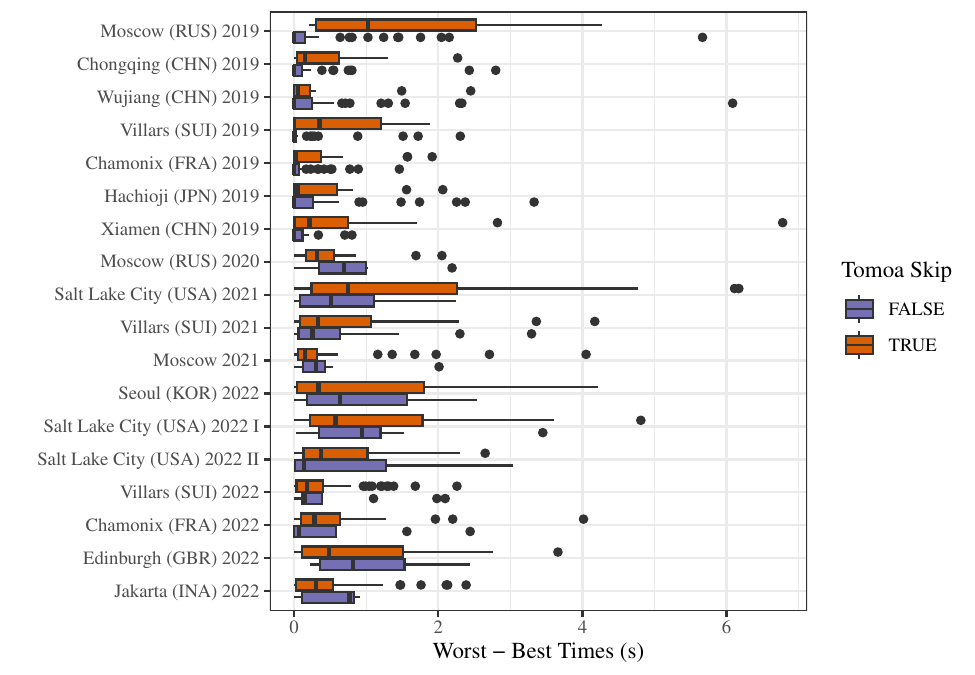} 

}

\caption{Boxplots of the ranges of climbers who use the Tomoa Skip vs those who do not. When the Tomoa Skip is first introduced there appears to be more variability in the ranges of Tomoa Skip users. This variability lessens as time goes on and becomes similar between those who use the Tomoa Skip and those who do not.}\label{fig:ts-ranges}
\end{figure}

Next, we can examine the ranges of climbing times for each climber in different events between those that use the Tomoa Skip and those who do not. The distribution of these ranges for each event after the introduction of the Tomoa Skip is shown in Figure \ref{fig:ts-ranges}. The range for an individual climber in an event is specified by their best time minus their worst time for that event. A climber with a low range can be said to be highly consistent in their climbing times. In 2019, when the Tomoa Skip is first being widely used in competitions, there seems to be more variability in the ranges of Tomoa Skip users when compared to those who did not use the Tomoa Skip. As time goes on the variability between Tomoa Skip users and non-Tomoa Skip users becomes increasingly similar. This could be the result of climbers becoming more familiar with the techniques required to successfully perform the Tomoa Skip as time goes on and climbers become more skilled with the strategy.

Another way to examine the relationship between the Tomoa Skip and variability of speed climbing times is to look at falls. Because the Tomoa Skip requires dynamic movement and strong technique it could be reasoned that it may be a riskier strategy for many speed climbers. In order to address this question we fit a generalized mixed effects model similar to the model used to analyze the effect of the Tomoa Skip on speed climbing times in Section \ref{variability}. This model differs from our previous model in that it uses falls as the outcome variable instead of the climbers best times and is generalized to the binomial distribution as no fell more than once in a competition. After fitting our model, we failed to identify a significant relationship between climbers falling and Tomoa Skip usage. Although visually we may see some patterns that indicate an increased variability in speed climbing times linked to Tomoa Skip usage, it appears that the Tomoa Skip does not have a statistically significant effect on the probability of a climber to fall in a competition.

\section{Final Thoughts}\label{final-thoughts}

\begin{figure}

{\centering \includegraphics[width=0.7\linewidth]{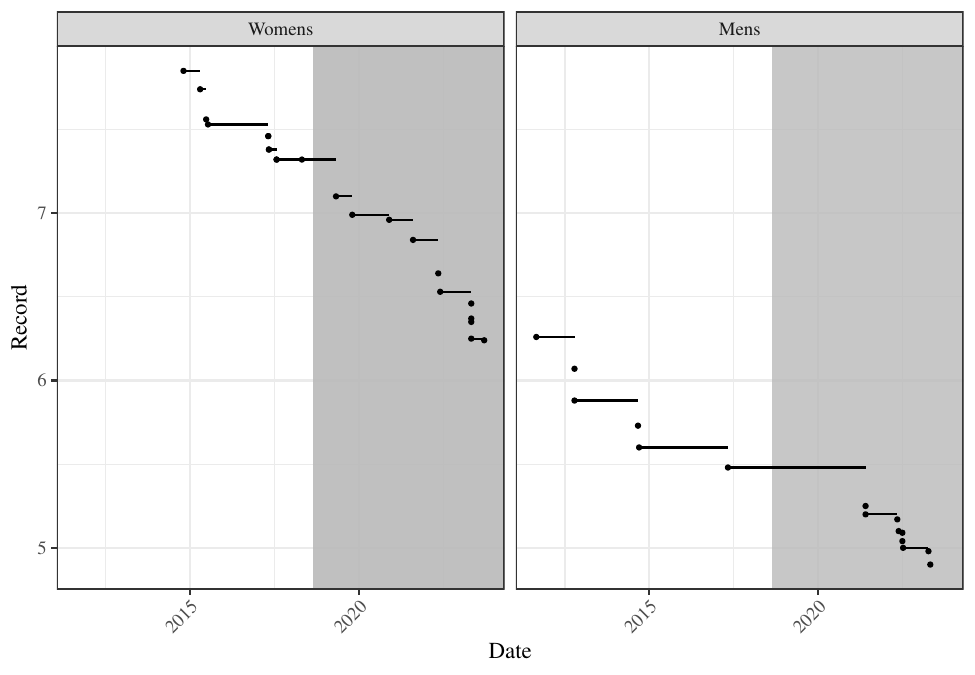} 

}

\caption{World record holdage over time. The grey rectangle denotes the period at which the Tomoa Skip has been introduced. After the Tomoa Skip is introduced, a slew of mens and womens records are broken over a short period of time.}\label{fig:wr-holdage}
\end{figure}

The introduction of the Tomoa Skip has imposed a notable impact on the speed climbing community. Between 2018 and 2023, the men's speed climbing record was broken a staggering nine times. The most recent record has pushed the sport into the sub 5 second zone for the first time in history. For reference, between the years 2012 and 2017, only six records were broken. A similar trend is also present in the recent storm of women's speed records with 12 records broken between 2018 and 2023 and seven records broken between 2013 and 2017. These patterns can be seen in Figure \ref{fig:wr-holdage}.

While interesting, our findings were somewhat limited by the data that was available. The absence of formally documented data on who has performed the Tomoa Skip limits our data to what we can view in recorded competitions, limiting our knowledge to climbers who have reached final rounds. This could potentially bias our findings towards those top speed climbers who are consistently reaching the final rounds of competitions. This fact also necessitated our assumption that those who use the Tomoa Skip once will continue to use it. Although our findings suggest that this is a reasonable assumption to make, we run the risk of overgeneralizing.

Although many were upset with the combined format of the 2020 Olympic games, it is not unreasonable to say that the groundbreaking and powerful Tomoa Skip would have never been introduced into the world of speed climbing if not for this format. The introduction of boulderer Tomoa Narasaki to speed climbing is what led to this unique convergence of disciplines where skills are shared and adapted between styles. The groundbreaking success of the Tomoa Skip highlights the potential of collaboration between different disciplines of sport, emphasizing that bringing together athletes of different backgrounds and styles may contribute to the overall health and evolution of competition.

\section*{Further Reading}\label{further-reading}
\addcontentsline{toc}{section}{Further Reading}

\phantomsection\label{refs}

\end{document}